\documentclass[12pt]{article}

\date{}

\begin{document}

\title{Eddington-Born-Infeld action and the dark side of general relativity.}

\author{M\'{a}ximo Ba\~{n}ados \\
P. Universidad Cat\'{o}lica de Chile,\\ Av. Vicu\~{n}a Mackenna 4860, Santiago, Chile,  \\ mbanados@uc.cl}

\maketitle

\begin{abstract}
We review a recent proposal to describe dark matter and dark energy based on an Eddington-Born-Infeld action. The theory is successful in describing the evolution of the expansion factor as well as galactic flat rotation curves. Fluctuations and the CMB spectra are currently under study. This paper in written in honor of Claudio Bunster on the occasion of his sixtieth birthday.
\end{abstract}

~

Marc Henneaux remarked in his Lecture at Claudio's Fest (Valdivia Chile, January 2008)  that working with him one learns  to be brave.  Claudio's tuition has been particularly important for me over the last two years. I have been working on an idea that looked crazy at first sight and still looks pretty mad today.   I have not had the chance to talk to Claudio about these ideas and I have not sent him the papers I have written \cite{B,B2} because I know he will not read them!  In this short contribution in his honor I attempt to shoot two birds with one stone, hoping that the bird will not be me.

I shall review a project whose aim is to provide a candidate for dark matter and dark energy, and whose seed relies on the study of general relativity around the degenerate field $g_{\mu\nu}=0$.  The project started as a purely formal idea, but it immediately succeeded in reproducing some of the known phenomenological curves associated to dark matter and dark energy.   I thus decided to purse the idea to the end.  In this contribution I shall concentrate on the original motivation based on studying general relativity near $g_{\mu\nu}=0$ \cite{B}.   This controversial motivation is now not needed because an action implementing most of the ideas is available \cite{B2}. However, I believe that exploring physics at $g_{\mu\nu}=0$ is an attractive idea, certainly not new, and perhaps necessary to understand the origin of the Universe.

The first step comes from the first order vielbein formulation of general relativity.
An intriguing solution to the equations of motion,
\begin{eqnarray}
  \epsilon_{abcd} R^{ab}e^c &=& 0 \label{e} \\
  \epsilon_{abcd} T^{a}e^c  &=& 0, \label{w}
\end{eqnarray}
is
\begin{equation}
e^a=0.
\end{equation}
The existence of this solution has not gone unnoticed, e.g., \cite{Witten,Horowitz,Giddings}. Its most salient property is that it preserves the full set of diffeomorphisms and for this reason it is often called the `unbroken' ground state of general relativity.  This point was stressed in \cite{Giddings} were a symmetry breaking transition from $e^a=0$ to $e^a\neq 0$, the Big-Bang, was suggested. Topological transitions in this formulation were studied in \cite{Horowitz}.

Now, a key aspect of this solution is the fact that the spin connection is left undetermined. The above equations of motion are supposed to determine both $a^a$ and $w^{ab}$. However, since $e^a=0$ kills both (\ref{e}) and (\ref{w}), the spin connection becomes a random field\footnote{This is related to another feature of $e^a=0$.  The leading term of the action $I[e,w] \sim \int \epsilon_{abcd}R^{ab} e^c e^d$ is cubic with respect to $w=e=0$, $\int\epsilon_{abcd}dw^{ab}e^ce^d$. Thus, around $e^a=0$ there is no quadratic term to expand, and no linearized theory can be defined.   Of course the action can be expanded around the `broken' solution $e^a_{\mu} = \delta^a_{\mu}$ with a well-defined linearized theory, but the interactions become non-renormalizable. In three  dimensions this problem does not occur because the action has one less power of $e^a$ and the quantum theory can be explored much further \cite{Witten}.}. The first step towards accepting $e^a=0$ as a solution is to understand the nature of the spin connection at that point.

If one first solves the algebraic equation for the torsion expressing $w\sim e^{-1}\partial e$ as a function of $e^a$ the same problem reappears in a different way.
One may try to recover the solution $e^a=0$ by a limit $e^a \rightarrow 0$. The connection $w \simeq e^{-1}\partial e$  has the structure ${0 \over 0}$ and can be anything\footnote{Note that in particular the limit may be a smooth differentiable function. In that case the curvature is well-defined. In particular $R_{\mu\nu}$ exists at the limit while $R=g^{\mu\nu}R_{\mu\nu}$ does not. Not surprisingly, metric invariants are not good objects to characterize the $g=0$ phase. }. This is equivalent to the statement that $w$ becomes random at $e^a=0$. One does learn something with this exercise though. The structure ${0 \over 0}$ appears provided both $e^a$ and its derivative vanish at all points.  This tells us that the limit $e^a\rightarrow 0$ cannot be  associated to time evolution (where $e^a=0$ would occur at some particular time $t_0$). Trying to understand the big-bang as a transition from zero metric into  non-zero metric raises complicated issues on the role of causality. Before the metric is created there is no causality at all\cite{Giddings}.

The problem we shall attack in this paper is the arbitrariness of the connection at $e^a=0$.  Our approach will consist on adding a new interaction to the action such that, as $e^a\rightarrow 0$, the spin connection does not go random but continues to be controlled by second order field equations.

Interestingly there are no too many terms one could add. For reasons we shall explain in a moment, it is necessary to go back to the metric formulation.    Consider the following Palatini action including a new term that depends only on the connection,
\begin{equation}\label{I0}
I[g,\Gamma] = \int \left[ \sqrt{g} (g^{\mu\nu}R_{\mu\nu}(\Gamma) + \Lambda)  + \kappa \sqrt{ |R_{\mu\nu}(\Gamma)|} \right].
\end{equation}
Here $R_{\mu\nu}$ is the Ricci tensor, which only depends on the connection, not the metric.  The new term is known as Eddington theory. The constant $\kappa$ is a coupling constant which in principle should be small enough such that this action is not in contradiction with well-known experiments. It is interesting to note the uniqueness of this term.  In the absence of a metric, Eddington's functional is the only density with the correct weight to respect diffeomorphism invariance. Note that Eddington's  action cannot be defined in the first order tetrad formalism. The $SO(3,1$ curvature $R^{a}_{\ b\mu\nu}(w)$ cannot be traced to produce a two index object, without using $e^a_\mu$. This is in contrast with the $GL(4)$ curvature $R^{\alpha}_{\ \beta\mu\nu }(\Gamma)$ whose trace $R_{\beta\nu}(\Gamma)$ is a tensor and independent from the metric.

The attractive feature of the action (\ref{I0}) is that if  the metric was not present, then the first two terms are not present and the dynamics is governed by Eddington's action\footnote{We borrow here the prescription from the tetrad formalism: $\epsilon_{abcd}R^{ab}e^c e^d \sim \sqrt{g}g^{\mu\nu}R_{\mu\nu}$ vanishes if $e^a \sim g_{\mu\nu} \rightarrow  0$. Another way to see this is by noticing that the volume element $\sqrt{g}$ scales faster than the metric inverse $g^{\mu\nu}$, at least for $d>2$.}. In this sense we have produced an action whose dynamics is well-defined even if the metric is switched off.

However, it is now a simple exercise to prove that the action (\ref{I0}) does not produce any interesting new effects.  Actually, this was already known to Eddington.   What happens is that the Einstein-Hilbert action with a cosmological term is dual to Eddington's action \cite{Fradkin}.  In other words, the Eddington term in (\ref{I0}) only renormalizes Newton's constant.

This can be seen as follows.
Consider the Palatini action for gravity with a cosmological constant,
\begin{equation}\label{IP}
I_P[g,\Gamma] = \int \sqrt{g} (g^{\mu\nu} R_{\mu\nu}(\Gamma) - 2\Lambda)
\end{equation}
It is well known that upon eliminating the connection using its own equation of motion one arrives at the usual second order Hilbert action
\begin{equation}\label{IH}
I_{H}[g] = \int \sqrt{g}(R(g)-2\Lambda).
\end{equation}
It is less well-known but also true \cite{Fradkin} that, if $\Lambda\neq 0$, then the metric can also be eliminated by using its own equations\footnote{This duality is of course well know to Claudio, and in fact the first time I heard about it was from him.}.   The variation ${\delta I \over \delta g^{\mu\nu}}=0$ yields,
\begin{equation}\label{IG}
g_{\mu\nu} = {1 \over \Lambda} R_{\mu\nu}(\Gamma).
\end{equation}
Since this is an algebraic relation for $g_{\mu\nu}$ it is legal to replace it back in the the action obtaining Eddington's functional
\begin{equation}\label{IE}
I_E[\Gamma]={2 \over \Lambda} \int \sqrt{\det(R_{\mu\nu})}.
\end{equation}
In the terminology of dualities, the action (\ref{IP}) is called the Parent action, while the Einstein-Hilbert action (\ref{IH}) and Eddington's action (\ref{IE}) are its   daughters. $I_H$ and $I_E$ are said to be dual to each other, and in many respects they are equivalent \cite{Fradkin,Boulanger}.

Summarizing, the action (\ref{I0}) can be understood as general relativity interacting with its own dual field. By a set of duality transformations we can transform the whole action (\ref{I0}) into standard general relativity with a new coupling constant. (Starting from (\ref{I0}) one eliminate the metric and to get Eddington action twice. Then apply a new transformation to get back to Einstein-Hilbert.)

An important note of caution is in order here. The equivalence between the Einstein-Hilbert and Eddington actions is true provided $g_{\mu\nu}$ is not degenerate. For degenerate fields they do represent different dynamics, and in fact, only Eddington's action is well-defined in that case.  The reason we shall not consider these cases is that, at the end of the day, we are interested in non-degenerate metrics anyway. Our guiding principle is to uncover what sort of modifications would be necessary to incorporate $g_{\mu\nu}=0$ as an allowed state. But, the physics phenomena we shall be interested do require a non-degenerate metric.

We shall now recall an analogy with condensed matter physics, suggesting a different interpretation for Eddington's action, which will truly depart from pure general relativity.

Within standard general relativity, we have observed that if the metric is removed then  the spin connection goes random. This looks very similar to a set of spins, at $T>T_c$, in the presence of an external magnetic field. As the field is removed the spins go random. However, if the temperature is below its critical value $T<T_c$, then the external field can be removed and the spins retain their ordered state. The crucial property of spins which makes this possible is their self interaction. Eddington's action has a similar role. For the action (\ref{I0}), as the metric is removed, the connection continues to be described by a well-defined set of equations\footnote{At this point we treat the metric or tetrad as an external field which can be switched on and off, as a mathematician would do. On a first approximation one does not look at the Maxwell equations governing the external field but simply assume that it can be controlled at will.  We have assumed the same with the metric, treating it as an external field. A full action governing the coupled system will be displayed below.}.

Now, spins also exhibit the opposite phenomena, namely spontaneous ``magnetization". If $T<T_c$ then the random disordered state is unstable and decays spontaneously into a broken ordered state. No external field is needed to trigger this phenomena. Let us imagine that the connection in general relativity exhibits a similar phenomena. That is, {\it without} introducing a metric, we assume that a connection can exists and be described by some well-defined equations. We shall call this connection $C^{\mu}_{\ \alpha\beta}$.  This field is fully independent from the metric.  Obviously, the only action consistent with general covariance is again Eddington's theory,
\begin{equation}
I[C] = \int \sqrt{|K_{\mu\nu}(C)|}
\end{equation}
where $K_{\mu\nu}(C)$ is the (traced) curvature  associated to the connection $C$. No metric is needed for this construction. What we have in mind is the existence of a connection $C^{\mu}_{\  \nu\alpha}$ that existed before the Universe, as a Riemannian     manifold, was created. This has to be interpreted with care because without $g_{\mu\nu}$ there are no causality relations.

At this point we have departed from the treatment suggested in \cite{Giddings}. In that reference the field which acquires an spontaneous non-zero expectation value was the metric itself.

We shall now turn on the metric and consider the whole self consistent system. The metric $g_{\mu\nu}$ generates its own connection, the Christoffel symbol $\Gamma(g)$. Thus, our theory will contain two independent connections. One, called $C$, is generated spontaneously. The other, called $\Gamma(g)$ is driven by the external field $g_{\mu\nu}$.

The action describing the coupled system is notably simple. The pair $g,\Gamma$ are of course described by the standard Einstein-Hilbert action either in first or second order form. On the other hand the field $C$ is described by Eddington's action.

Now, to make things more interesting we shall couple both connections through the metric. Again, there are no two many couplings one can write. An attractive possibility  is the Einstein-Hilbert-Eddington-Born-Infeld action \cite{B2},
\begin{equation}\label{I}
I = \int \sqrt{g}(R-2\Lambda) + {2 \over \alpha l^2}\sqrt{|l^2K_{\mu\nu} - g_{\mu\nu}|}.
\end{equation}
This action has several interesting properties. First,
note that as $g\rightarrow 0$ we recover Eddington's action for the field $C$. (We shall not need this limit in what follows.) Second, (\ref{I}) is a tensor Born-Infeld theory analogous to the scalar $\sqrt{\det (g_{\mu\nu}+ \partial_\mu\phi\partial_\nu\phi)}$ and vector  $\sqrt{\det (g_{\mu\nu}+ F_{\mu\nu})}$ Born-Infeld theories. Observe that the equations of motion for the whole action are of second order. This is different from the gravitational BI action written in \cite{Deser-G} where extra terms had to be added in order to eliminate the ghost. Finally, the action (\ref{I}) contains dynamical properties which makes it an attractive candidate for dark matter and dark energy\cite{B2}. Both appear in a unified way, just like the Chapligyn formulation \cite{Chapligyn}. It is also curious to observe that the Chapligyn gas can be derived from a scalar Born-Infeld theory.

More specifically, for Friedman models, it follows that the Eddington field $C$ behaves like matter for early times and as dark energy for late times: its equation of state $w=p/\rho$ evolve from $w=0$ near the big bang to $w=-1$ for late times. One can also analyze the dynamics of objects moving around spherically symmetric sources. The Eddington field in this case yields asymptotically flat rotation curves and thus again provides a candidate for dark matter. More details on these issues can be found in \cite{B2}.

The reader may wonder what does $g_{\mu\nu}=0$ have to do with dark matter and dark energy. Could one had predicted this (suggested) relationship?   Our initial motivation to look at $g_{\mu\nu}=0$ came from the following analogy. A particle at rest has an energy $mc^2$.  The most direct manifestation of this energy is through gravity,  and in fact $mc^2$ is a source of curvature.  From a Newtonian point of view, the energy of a particle at rest is zero.  Can we ask the same question in general relativity? Could flat space have an energy-density associated to it?   In the standard choice for zero point of energy, flat space has zero energy. The very definition of energy in general relativity requires knowledge of boundary conditions and certainly $g_{\mu\nu}=0$ falls outside all examples. However at the level of energy density, namely, the Einstein tensor $G_{\mu\nu}(g)$, one may wonder about its value at $g_{\mu\nu}=0$. Interestingly this tensor depends only on $g^{-1}\partial g$ and thus the limit can be defined in a way that $G_{\mu\nu}(0)$ becomes finite. By a mixture of Bianchi identities and some reasonable assumptions its value can be computed and yields a contribution to Einstein equations similar to those expected from dark matter \cite{B}.

The ideas presented in this contribution are highly  speculative and need formalization. The analogy with spin systems is the most challenging and difficult problem. Other  applications like fluctuations  and the CMB spectra are presently under analysis \cite{BFS}.  The action (\ref{I}) has several interesting formal properties under duality transformations. This theory can be written as a bigravity theory, which have been under great scrutiny in the past and also recently \cite{Hamed}. A detailed analysis will be reported in \cite{BGR}. Rotation curves for several galaxies has been analyzed in \cite{RR}, with interesting results.

~

~

Some of the ideas presented here have been developed in collaboration with A. Gomberoff and D. Rodriguez. I would like to thank them for their permission to include here some unpublished material.  I would also like to thank S. Theisen for many useful conversations and for bringing to my attention Ref. \cite{Fradkin}. P. Ferreira and C. Skordis have been crucial to keep this idea alive.

\end{document}